\documentclass[letter]{aa} 
\bibpunct{(}{)}{;}{a}{}{,}
\usepackage{graphicx}
\usepackage{natbib}
\usepackage[toc,page]{appendix}
\usepackage{txfonts}
\usepackage{multicol}
\usepackage{amsmath}
\usepackage{hyperref}
\usepackage{longtable}
\usepackage{lscape}
\begin{document} 
\title{The CARMENES search for exoplanets around M dwarfs}
\subtitle{No evidence for a super-Earth in a 2-day orbit around GJ\,1151}
\author{M.~Perger\inst{1,2}
		\and I.~Ribas\inst{1,2}
		\and G.~Anglada-Escud\'e\inst{1,2}
		\and J.~C.~Morales\inst{1,2}		
		\and P.\,J.~Amado\inst{3}
		\and J.\,A.~Caballero\inst{4}
		\and A.~Quirrenbach\inst{5}
		\and A.~Reiners\inst{6} 
		\and V.\,J.\,S.~B\'ejar\inst{7,8}	
		\and S.~Dreizler\inst{6}
		\and D.~Galad\'i-Enr\'iquez\inst{9}  
		\and A.\,P.~Hatzes\inst{10}
		\and Th.~Henning\inst{11}
		\and S.\,V.~Jeffers\inst{12}
		\and A.~Kaminski\inst{5} 
		\and M.~K{\"u}rster\inst{11}
		\and M.~Lafarga\inst{1,2,13}
		\and D.~Montes\inst{14}
		\and E.~Pall\'e\inst{7,8}		
		\and C.~Rodr\'iguez-L\'opez\inst{3} 
		\and A.~Schweitzer\inst{15}
		\and M.~R.~Zapatero Osorio\inst{16} 
		\and M.~Zechmeister\inst{6}	
}
\offprints{M.~Perger,
\email{perger@ice.cat}}
\institute{
		\inst{1}Institut de Ci\`encies de l'Espai (ICE, CSIC), Campus UAB, Carrer de Can Magrans s/n, 08193 Bellaterra, Spain\\
		\inst{2}Institut d'Estudis Espacials de Catalunya (IEEC), 08034 Barcelona, Spain\\
		\inst{3}Instituto de Astrof\'{\i}sica de Andaluc\'{\i}a (IAA-CSIC), Glorieta de la Astronom\'{\i}a s/n, 18008 Granada, Spain\\
        \inst{4}Centro de Astrobiolog\'{\i}a (CSIC-INTA), ESAC, Camino bajo del castillo s/n, 28692 Villanueva de la Ca\~nada, Madrid, Spain\\
		\inst{5}Landessternwarte, Zentrum f\"ur Astronomie der Universit\"at Heidelberg, K\"onigstuhl 12, 69117 Heidelberg, Germany\\
		\inst{6}Institut f\"ur Astrophysik, Georg-August-Universit\"at G\"ottingen, Friedrich-Hund-Platz 1, 37077 G\"ottingen, Germany\\	
        \inst{7}Instituto de Astrof\'{\i}sica de Canarias (IAC), 38205 La Laguna, Tenerife, Spain\\
        \inst{8}Departamento de Astrof\'{\i}sica, Universidad de La Laguna (ULL), 38206, La Laguna, Tenerife, Spain\\
        \inst{9}Centro Astron\'omico Hispano-Alem\'an (CAHA), Observatorio de Calar Alto, Sierra de los Filabres, 04550 G\'ergal, Spain\\
        \inst{10}Th\"uringen Landessternwarte Tautenburg, Sternwarte 5, D-07778 Tautenburg, Germany\\
        \inst{11}Max Planck Institute for Astronomy, K\"onigstuhl 17, 69117 Heidelberg, Germany\\
        \inst{12}Max Planck Institute for Solar System Research, Justus-von-Liebig-Weg 3, 37077 G\"ottingen, Germany\\
        \inst{13}Department of Physics, University of Warwick, Gibbet Hill Road, Coventry CV4 7AL, UK\\
        \inst{14}Departamento de F\'{\i}sica de la Tierra y Astrof\'{\i}sica \& IPARCOS-UCM (Instituto de F\'{\i}sica de Part\'{\i}culas y del Cosmos de la UCM), Facultad de Ciencias F\'{\i}sicas, Universidad Complutense de Madrid, 28040 Madrid, Spain\\
        \inst{15}Hamburger Sternwarte, Gojenbergsweg 112, 21029 Hamburg, Germany\\
		\inst{16}Centro de Astrobiolog\'{\i}a (CSIC-INTA), Carretera de Ajalvir km 4, 28850 Torrej\'on de Ardoz, Madrid, Spain
	} 
\date{Received: 11 March 2021/ Accepted: dd Month 2021}
\abstract
	{The interaction between Earth-like exoplanets and the magnetic field of low-mass host stars are considered to produce weak emission signals at radio frequencies. A study using LOFAR data announced the detection of radio emission from the mid M-type dwarf GJ\,1151 that could potentially arise from a close-in terrestrial planet. Recently, the presence of a 2.5-M$_{\oplus}$ planet orbiting GJ\,1151 with a 2-day period has been claimed using 69 radial velocities (RVs) from the HARPS-N and HPF instruments.}
	{We have obtained 70 new high-precision RV measurements in the framework of the CARMENES M-dwarf survey and use these data to confirm the presence of the claimed planet and to place limits on possible planetary companions in the GJ\,1151 system.}
	{We analyse the periodicities present in the combined RV data sets from all three instruments and calculate the detection limits for potential planets in short-period orbits.}
	{We cannot confirm the recently-announced candidate planet and conclude that the 2-day signal in the HARPS-N and HPF data sets is most probably produced by a long-term RV variability possibly arising from a yet unconstrained outer planetary companion. We calculate a 99.9\% significance detection limit of $1.50$\,m\,s$^{-1}$ in the RV semi-amplitude, which places upper limits of 0.7\,M$_{\oplus}$ and 1.2\,M$_{\oplus}$ to the minimum masses of potential exoplanets with orbital periods of 1 and 5 days, respectively.}
	{}
\keywords{techniques: spectroscopic -- stars: late-type -- stars: planetary systems -- stars: individual: GJ\,1151}
\maketitle
%
	
\section{Introduction}
	
Recently, \citet{2020NatAs...4..577V} reported the detection of circularly-polarised radio emission in the LOFAR \citep[LOw-Frequency ARray;][]{2013A&A...556A...2V} Two-Metre Sky Survey (LoTSS) data release I \citep{2019A&A...622A...1S}. It was detected over a relatively long duration ($>8$\,h), and at low frequency ($\sim$150\,MHz), at the position in the sky of the M4.5-type \citep{2013AJ....145..102L} star GJ\,1151, for which we calculated a mass of $0.170\pm0.010\,$M$_{\odot}$ \citep{2019A&A...625A..68S}. The star is a slow rotator with $v \sin{i} < 2$\,km\,s$^{-1}$ \citep{2018A&A...612A..49R}, and with a photometric rotation period estimated at $P_{\rm rot}=117.6$\,d \citep{2016ApJ...821...93N} and $P_{\rm rot}=125\pm23$\,d \citep{2019A&A...621A.126D}. All available evidence, including a measured pseudo-equivalent width of the H$\alpha$ line pEW(H$\alpha$)=$+0.342\pm0.008$\,\AA~\citep[following][]{2019A&A...623A..44S}, points at a very low magnetic activity in the star.
	
\citet{2020NatAs...4..577V} discuss that the Poynting flux required to produce the detected radio signal cannot be generated by a star with such characteristics and hence suggest that its origin is rather related to the interaction with a companion. The possibility for it to be a long-period substellar massive object was already ruled out by FastCam observations \citep{2017A&A...597A..47C}, at least for separations $>1$\,au. Thus, the authors suggest the existence of a short-period ($P$=1--5\,d) Earth-like planet with an orbit interior to the habitable zone of the star. Then, they argue, the radio signal could originate from the sub-Alfv\'enic interaction of this planet with the plasma of the stellar magnetosphere inducing electron cyclotron maser instability \citep{1982ApJ...259..844M}. Recent results of {\em XMM-Newton} X-ray data \citep{2020MNRAS.497.1015F} seem to strengthen this assumption. Since this effect is expected to be very weak, the detection of an exoplanet at radio wavelengths is very intriguing.
	
\begin{table*}
	\caption{\label{T1} Statistics of the different RV data sets.}
	\centering
	\begin{tabular}{lccccccc}
		\hline \hline \noalign{\smallskip}
		Instrument & Unit & HARPS-N & HARPS-N & HPF & CARMENES & Combined & Combined\\
		RV extraction code	& & {\sc terra} & {\tt wobble} & {\sc serval}  & {\sc serval} & ... & residuals \\
		\noalign{\smallskip}
		\hline
		\noalign{\smallskip}
		$N_{\rm obs}$ & ... & 20 & 19 & 25 & 70 & 115 & 115  \\
		rms & [m\,s$^{-1}$] & 2.61 & 3.63 & 4.64 & 4.09 & 4.15 & 3.32\\
		$\delta RV$ & [m\,s$^{-1}$]   & 1.85 & 2.88 & 3.00 & 1.79 & 2.06 & 2.06 \\ 
		$T$ & [d]  &  69 & 69 & 468  & 1\,793 & 1\,793 & 1\,793 \\
		$\Delta t$ & [d]  & 3.6 & 3.8 & 19.5 & 26.0 & 15.7 & 15.7\\
		\noalign{\smallskip} \hline 
	\end{tabular}
	\tablefoot{$N_{\rm obs}$ is the number of observations, rms is the RV root mean square, $\delta RV$ the mean RV uncertainty, $T$ the time baseline of the observations, and $\Delta t$ the median time sampling between epochs.}		
\end{table*}

The existence of such a planet was initially evaluated by \citet{2020ApJ...890L..19P} using 19 epochs of HARPS-N \citep[High Accuracy Radial velocity Planet Searcher of the Northern hemisphere;][]{2012SPIE.8446E..1VC} radial velocity (RV) data. The authors did not find any significant signal but placed an upper limit of $M\,\sin{i}<5.6\,{\rm M}_{\oplus}$ on the minimum mass of any possible close-in planet, assuming a stellar mass of $0.167\pm0.025\,{\rm M}_{\odot}$ \citep{2016ApJ...821...93N} and conclusively ruled out close-in stellar or gas-giant companions. More recently, \citet{2021arXiv210202233M} analysed the same HARPS-N RVs together with 50 epochs of newly obtained HPF \citep[Habitable-zone Planet Finder;][]{2012SPIE.8446E..1SM} near-infrared RVs. The authors reported a significant Doppler signal compatible with an $M\,\sin{i}=2.5\pm0.5\,{\rm M}_{\oplus}$ planet on a 2.02-day orbit inducing an RV semi-amplitude of $K = 4.1\pm0.8$\,m\,s$^{-1}$. 

Here, we report on the combined analysis of the published HARPS-N and HPF RVs together with an additional data set consisting of 70 epochs of CARMENES \citep[Calar Alto high-Resolution search for M dwarfs with Exoearths with Near-infrared and optical Echelle Spectrographs;][]{2020SPIE11447E..3CQ} RVs of GJ\,1151.
	
\section{Radial velocity analysis}
	
\begin{figure}
	\resizebox{\hsize}{!}{\includegraphics[width=\textwidth]{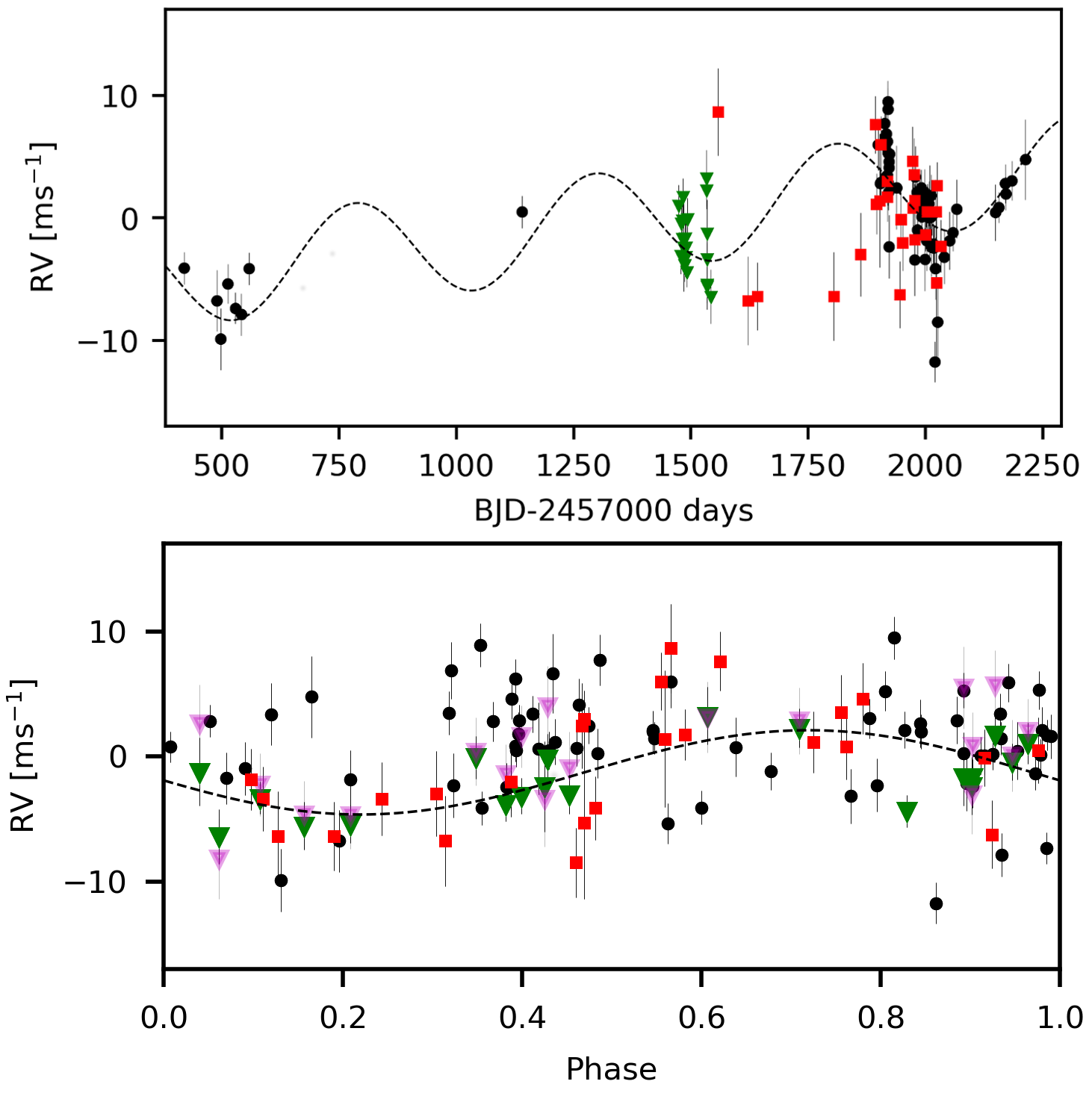}}
	\caption{RV data as observed from HARPS-N (green triangles), HPF (red squares), and CARMENES (black dots) in time series ({\it top panel}), and phase-folded ({\it bottom panel}) to the 2.02-day period of the planet candidate announced by \citet{2021arXiv210202233M}. In the top panel, we include the best-fitting linear trend of 1.73\,m\,s$^{-1}$ and long-period signal of $\sim 500$\,d represented by the black dashed line. In the bottom panel, we show additionally the HARPS-N data as derived by the {\tt wobble} code represented by magenta triangles.}
	\centering
	\label{F1a}	
\end{figure}

We used the 20 publicly available HARPS-N spectra of GJ\,1151 and calculated RVs with the {\sc terra} pipeline \citep{2012ApJS..200...15A}. The measurements show significantly smaller variations and uncertainties than the 19 RVs (see Table\,\ref{T1}) from both \citet{2020ApJ...890L..19P} and \cite{2021arXiv210202233M}, which were derived with the {\tt wobble} code \citep{2019AJ....158..164B}. The observations were acquired from December 2018 to February 2019, with occasional dense sampling (top panel of Fig.\,\ref{F1a}). In addition, Mahadevan (priv. comm.) kindly provided us with 25 RVs from the HPF near-infrared observations obtained between March 2019 and June 2020 and computed with the {\sc serval} code \citep{2018A&A...609A..12Z}. These RVs correspond to nightly averages of the 50 individual measurements presented in their paper. The data points show significantly larger individual uncertainties than the RVs from the HARPS-N instrument.
	
\begin{figure}		
	\centering
	\resizebox{!}{!}{\includegraphics[width=8.6cm]{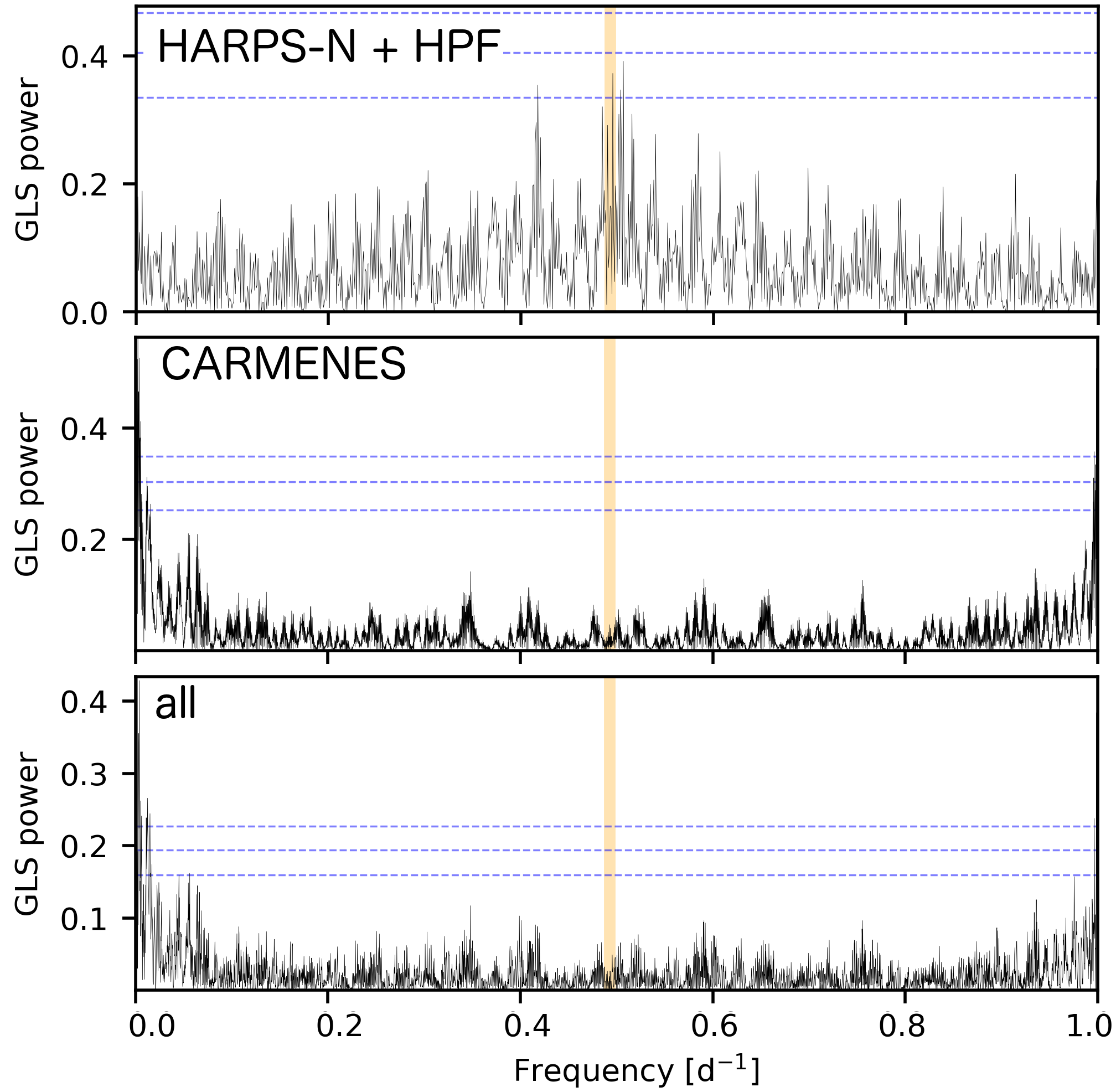}}
	\caption{Periodograms of the RVs as observed of the combined HARPS-N and HPF data sets ({\it top panel}), the individual CARMENES data set ({\it middle panel}), and the combined full RV data set ({\it bottom panel}). The orange vertical line highlights a period of 2.02\,d, and the horizontal blue dashed lines indicate analytical false-alarm probabilities of 10, 1, and 0.1\% (from bottom to top). The y axis is shown up to the largest GLS power or the 0.1\% FAP level.}
	\centering
	\label{F1d}	
\end{figure}
	
GJ\,1151 was observed on 70 occasions from 2016 to 2020 with CARMENES. The RVs of the visible channel were extracted with the {\sc serval} code (for statistics see Table\,\ref{T1}). The RVs of the NIR channel show average uncertainties of $\sim8$\,m\,s$^{-1}$ for this star and were not used in this study. The data are separated in two blocks, one of 7 measurements from February to June 2016 and one of 62 measurements, with more intensive monitoring periods, from February to December 2020 (top panel of Fig.\,\ref{F1a}). There is also one single measurement in-between the two blocks. An apparent global upward trend is clearly visible, and also some modulation with a period $>300$\,d. The data do not overlap with the HARPS-N measurements, but they do with most of the HPF data. Both sets clearly show a quite steep downward trend around BJD=2\,458\,900\,d.
	
As an initial test, we phase-folded the full dataset to the 2.02-d period of the announced planet candidate by fitting individual offsets for each instrument and using the best-fit semi-amplitude for the RVs of the present work. This is graphically shown in the bottom panel of Fig.\,\ref{F1a}. While the HPF data seem to favour such a fit, therefore reproducing the results of \cite{2021arXiv210202233M}, the HARPS-N data show phases with poor coverage, which then make the data compatible with the Keplerian signal found by the HPF data. In clear contrast, the CARMENES data do not confirm the modulation nor show any sign of periodic variability at 2.02\,d. This can also be seen in the periodograms of Fig.\,\ref{F1d} for the combined HARPS-N and HPF sets (top panel), the individual CARMENES RVs (middle panel) and the full data set (bottom panel).

\begin{figure}
	\resizebox{\hsize}{!}{\includegraphics[width=\textwidth]{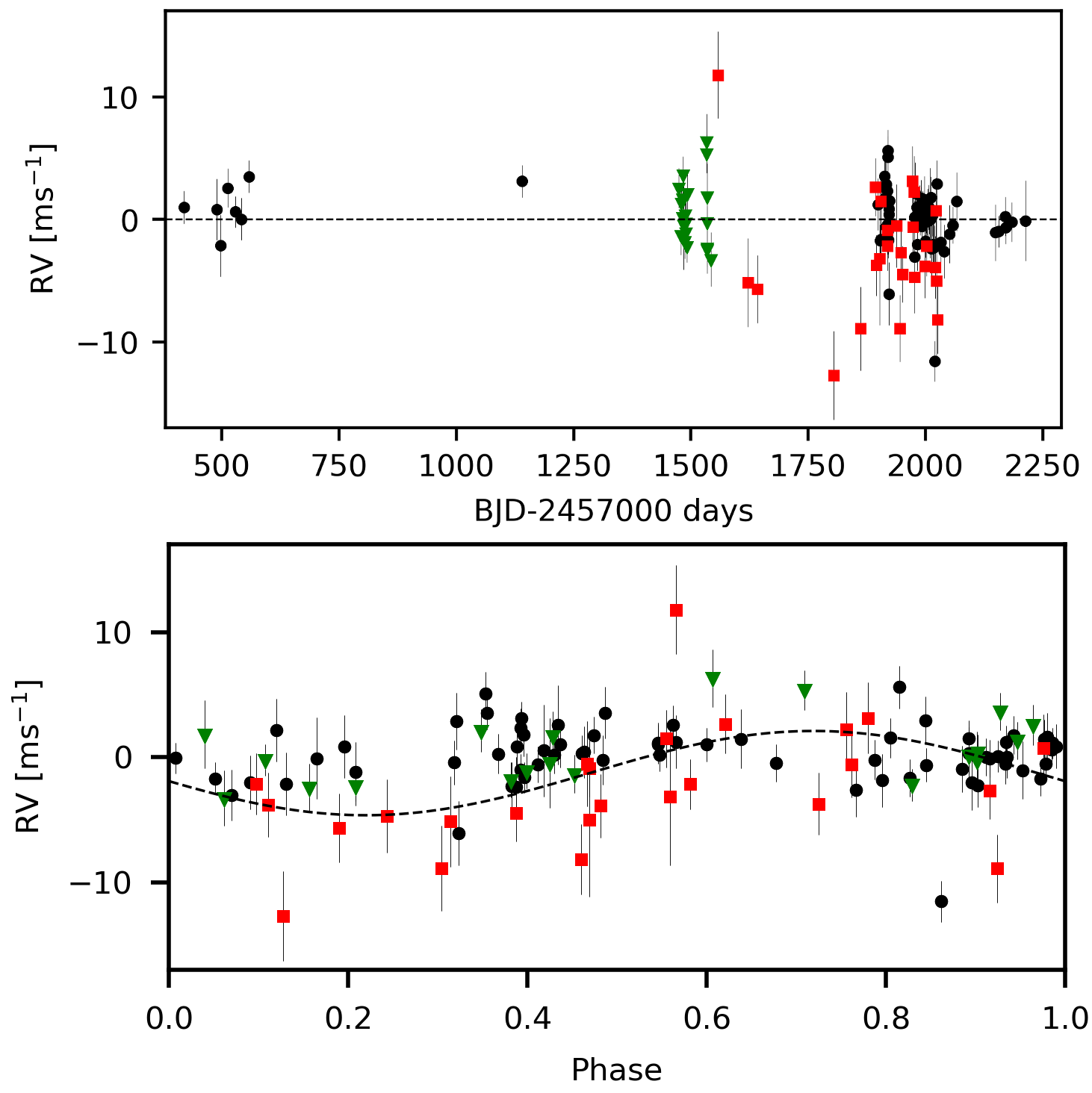}}
	\caption{Same as Fig.\,\ref{F1a}, but for the RV residuals that result after subtracting a linear trend and a long-period signal.}
	\centering
	\label{F1b}	
\end{figure}

\begin{figure}		
	\centering
	\resizebox{\hsize}{!}{\includegraphics[width=\textwidth]{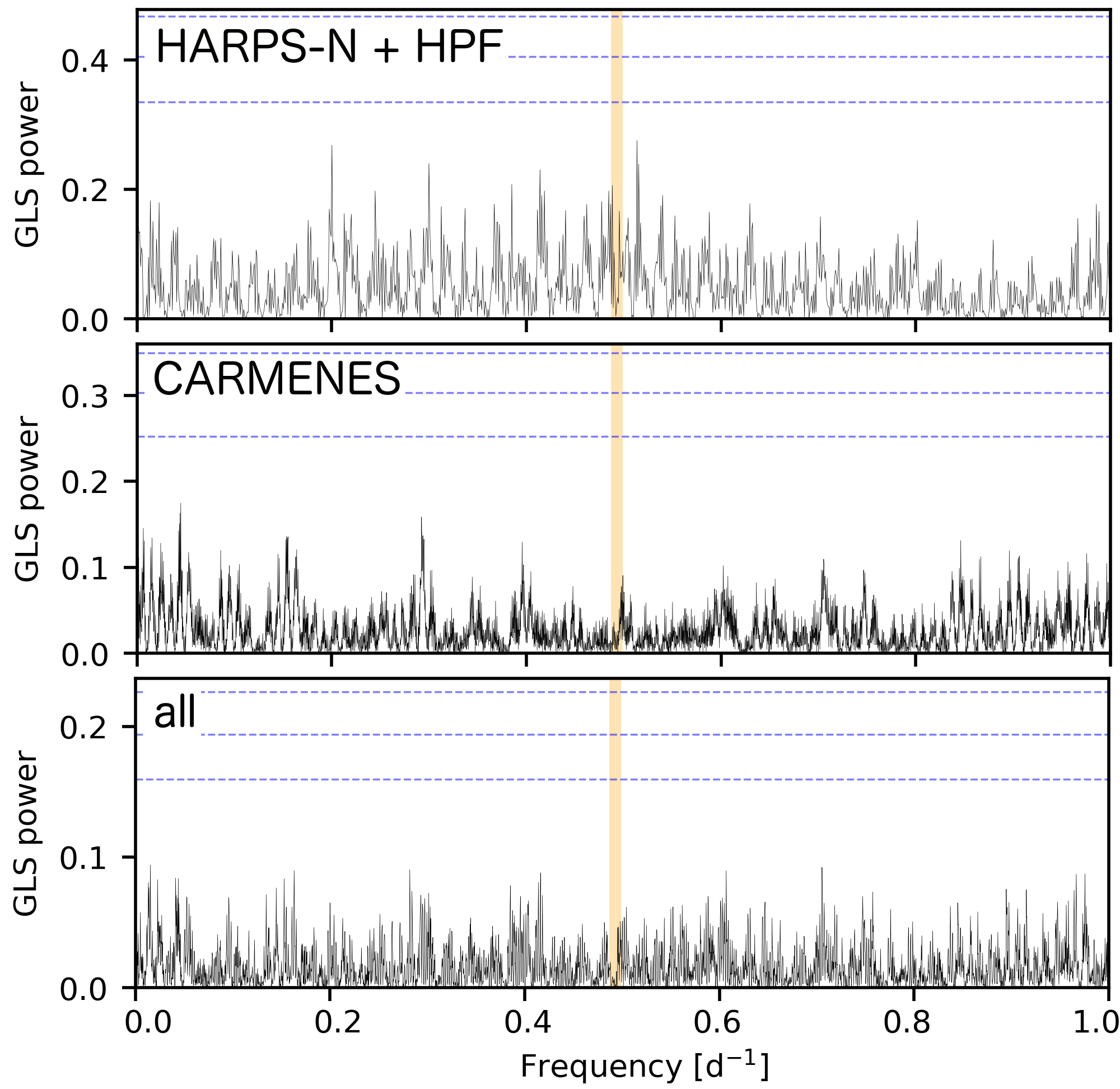}}
	\caption{Same as Fig.\,\ref{F1d}, but for the RV residuals that result after subtracting a linear trend and a long-period signal.}
	\centering
	\label{F1c}	
\end{figure}

The combined RV time series of the three instruments suggests the presence of a linear trend and a long-term modulation. Thus, we considered these two effects and optimised (maximum likelihood) their parameters together with the RV offsets amongst the different data sets. As a result, we find a linear trend of 1.73\,m\,s$^{-1}$\,yr$^{-1}$ and a highly significant signal with a period $\sim500$\,d and a semi-amplitude of $K=4.2$\,m\,s$^{-1}$ in the combined data.  Because of seasonal gaps, this long-period signal is not fully sampled in phase. The combined fit of the trend and the long-term signal is shown in the top panel of Fig.\,\ref{F1a}. The residuals of subtracting this fit from the data are shown as a time series in the top panel of Fig.\,\ref{F1b}, and folded to the 2.02-day period of the candidate planet in the bottom panel of Fig.\,\ref{F1b}. It is readily seen that none of the individual data sets, including HARPS-N and HPF, longer supports the existence of a significant periodicity. This is also observed in the periodograms of the residuals in Fig.\,\ref{F1c} of the combined HARPS-N and HPF data sets (top panel), the individual CARMENES data set (middle panel), and the combined full RV data set (bottom panel). No prominent periodic signals are visible, including a lack of significant periodicity at 2.02\,d. The signal identified in the HPF data is removed by the fit of the linear trend and the long-period signal. We therefore conclude that it was most likely a spurious signal caused by the dominant downward trend of the HPF data during the densely sampled epoch.

\begin{figure}
	\resizebox{\hsize}{!}{\includegraphics[width=\textwidth]{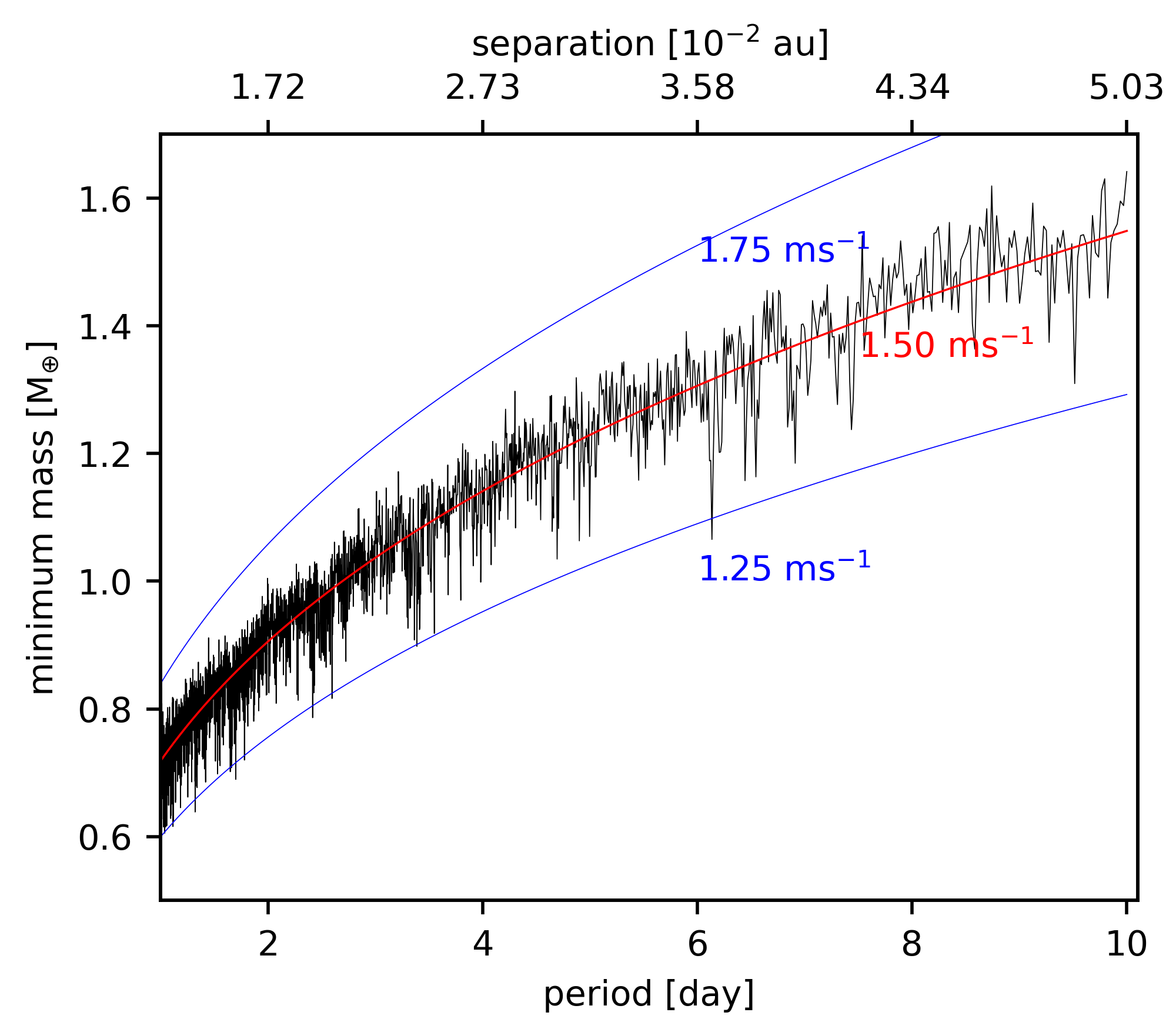}}
	\caption{Detection limits of the RV residuals (after correcting for a trend and long-term modulation) of the combined RVs of GJ\,1151 of 20 HARPS-N, 25 HPF, and 70 CARMENES observations. We show the minimum planetary mass for which we detect (bootstrap false-alarm probability $<0.1$\,\%) an injected planetary companion for each period (black line). Coloured lines show constant semi-amplitudes of 1.25, and 1.75\,m\,s$^{-1}$ (blue lines), and the average detection limit of 1.50\,m\,s$^{-1}$ (red line).}
	\centering
	\label{F2}
\end{figure}

We follow the procedure described in \citet{2013A&A...549A.109B} to calculate the detection limits for the RV dataset and the limit for the minimum masses of planets with 1- to 5-d orbital periods. We employ a significance threshold at a false-alarm probability of 0.1\,\%. We firstly considered the RV time series as observed, that is, without subtracting the trend and long-term modulation, and we obtained a flat detection limit of $K=2.21\pm0.15$\,m\,s$^{-1}$ for the RV semi-amplitude of circular orbits with periods between 1 and 5\,d. When we run the same calculations on the residuals after performing the correction, we derive a mean limiting RV semi-amplitude of $K=1.50\pm0.07$\,m\,s$^{-1}$, which translates into minimum planet masses of 0.72, 0.91, and 1.23\,M$_{\oplus}$ for orbital periods of 1, 2.02, and 5 days, respectively. A graphical representation of the experiment is shown in Fig.\,\ref{F2}. Since the scatter of the RV residuals (Table\,\ref{T1}) is of the order of 3.3\,m\,s$^{-1}$, the simulations show that we would be able to detect planets with semi-amplitudes some 2.2 times smaller than such velocity scatter. 

\section{Conclusions}

We analysed published HARPS-N and HPF RVs of the low-mass star GJ\,1151 together with 70 new CARMENES RVs, following up on the recent announcement of a possible planet being responsible for low-frequency radio emission detected by LOFAR. The full combined data set shows a linear trend of 1.73\,m\,s$^{-1}$\,yr$^{-1}$ and contains a long-period signal $>300$\,d. We are not yet able to unambiguously derive the parameters and to assess the nature of the suggestive, potentially planetary, long-period signal, but observations are still ongoing and will be investigated in an upcoming article.
	
If we subtract a trend and a long-period signal from the observations, effectively applying a high-pass frequency filter, the resulting residual RVs show no signs of the 2.5~M$_{\oplus}$ planet in a 2.02-day orbit proposed by \cite{2021arXiv210202233M}, which would induce a periodic RV signal with a semi-amplitude of 4.1\,m\,s$^{-1}$. We find that the reported periodic signal may rather be produced by the unaccounted-for long-period signal and the free offset used when combining both HPF and HARPS-N datasets. 
	
In our study of the full RV data, we place an new upper limit to the semi-amplitude of a possible exoplanet orbiting GJ\,1151 at 1.50\,m\,s$^{-1}$. A putative planetary companion with an orbit below 5\,days, as put forward to explain the LOFAR data, would need to have a minimum mass lower than 1.2\,M$_{\oplus}$ to remain compatible with the available RV dataset.
	
\begin{acknowledgements}

CARMENES is an instrument at the Centro Astron\'omico Hispano-Alem\'an (CAHA) at Calar Alto (Almer\'{\i}a, Spain), operated jointly by the Junta de Andaluc\'ia and the Instituto de Astrof\'isica de Andaluc\'ia (CSIC). The authors wish to express their sincere thanks to all members of the Calar Alto staff for their expert support of the instrument and telescope operation. CARMENES was funded by the Max-Planck-Gesellschaft (MPG), the Consejo Superior de Investigaciones Cient\'{\i}ficas (CSIC), the Ministerio de Econom\'ia y Competitividad (MINECO) and the European Regional Development Fund (ERDF) through projects FICTS-2011-02, ICTS-2017-07-CAHA-4, and CAHA16-CE-3978, and the members of the CARMENES Consortium (Max-Planck-Institut f\"ur Astronomie, Instituto de Astrof\'{\i}sica de Andaluc\'{\i}a, Landessternwarte K\"onigstuhl, Institut de Ci\`encies de l'Espai, Institut f\"ur Astrophysik G\"ottingen, Universidad Complutense de Madrid, Th\"uringer Landessternwarte Tautenburg, Instituto de Astrof\'{\i}sica de Canarias, Hamburger Sternwarte, Centro de Astrobiolog\'{\i}a and Centro Astron\'omico Hispano-Alem\'an), with additional contributions by the MINECO, the Deutsche Forschungsgemeinschaft through the Major Research Instrumentation Programme and Research Unit FOR2544 ``Blue Planets around Red Stars'', the Klaus Tschira Stiftung, the states of Baden-W\"urttemberg and Niedersachsen, and by the Junta de Andaluc\'{\i}a. This work was based on data from the CARMENES data archive at CAB (CSIC-INTA). We acknowledge financial support from the Agencia Estatal de Investigaci\'on of the Ministerio de Ciencia, Innovaci\'on y Universidades and the ERDF through projects PID2019-109522GB-C5[1:4]/AEI/10.13039/501100011033, PGC2018-098153-B-C33, AYA2018-84089, ESP2017-87676-C5-1-R, and the Centre of Excellence ``Severo Ochoa'' and ``Mar\'ia de Maeztu'' awards to the Instituto de Astrof\'isica de Canarias (SEV-2015-0548), Instituto de Astrof\'isica de Andaluc\'ia (SEV-2017-0709), and Centro de Astrobiolog\'ia (MDM-2017-0737), and the Generalitat de Catalunya/CERCA programme.

\end{acknowledgements}
	
\bibliographystyle{aa} 
\bibliography{bibtex}{}

\end{document}